\documentclass[aip,apl,reprint,superscriptaddress,floatfix,balancelastpage,UTF8]{revtex4-2}
\usepackage{amsmath}
\usepackage{mathptmx} 
\usepackage{graphicx}
\usepackage{fancyhdr}
\usepackage{ifthen}
\usepackage{multirow}
\usepackage{footmisc}
\usepackage{array}
\usepackage[colorlinks,
linkcolor=blue,       
anchorcolor=blue,  
citecolor=blue,
urlcolor=blue,
]{hyperref}
\newcommand{\setParDis}{\setlength{\parskip}{0cm}}
\usepackage[capitalize]{cleveref}

\usepackage{lipsum}

\DeclareSymbolFont{myletters}{OML}{ztmcm}{m}{it}
\DeclareMathSymbol{\uplambda}{\mathord}{myletters}{"15}

\begin{document}

	\title{Vertical Ferroelectricity in Van der Waals Materials: Models and Devices}

	\author{Yuwen Zhang}

	\date{\today}
	\affiliation{Hunan Key Laboratory for Micro-Nano Energy Materials and Device and School of Physics and Optoelectronics, Xiangtan University, Xiangtan 411105, Hunan, China}

	\author{Chunfeng Cui}

	\affiliation{Hunan Key Laboratory for Micro-Nano Energy Materials and Device and School of Physics and Optoelectronics, Xiangtan University, Xiangtan 411105, Hunan, China}

	\author{Chaoyu He}
\thanks{Authors to whom correspondence should be addressed: Chaoyu He, \textcolor{blue}{chaoyu@xtu.edu.cn}; Tao Ouyang, \textcolor{blue}{ouyangtao@xtu.edu.cn}; Chao Tang, \textcolor{blue}{tang$\_$chao@xtu.edu.cn.} }
	\affiliation{Hunan Key Laboratory for Micro-Nano Energy Materials and Device and School of Physics and Optoelectronics, Xiangtan University, Xiangtan 411105, Hunan, China}

	\author{Tao Ouyang}
\thanks{Authors to whom correspondence should be addressed: Chaoyu He, \textcolor{blue}{chaoyu@xtu.edu.cn}; Tao Ouyang, \textcolor{blue}{ouyangtao@xtu.edu.cn}; Chao Tang, \textcolor{blue}{tang$\_$chao@xtu.edu.cn.} }
	\affiliation{Hunan Key Laboratory for Micro-Nano Energy Materials and Device and School of Physics and Optoelectronics, Xiangtan University, Xiangtan 411105, Hunan, China}

	\author{Jin Li}
	\affiliation{Hunan Key Laboratory for Micro-Nano Energy Materials and Device and School of Physics and Optoelectronics, Xiangtan University, Xiangtan 411105, Hunan, China}
\author{\textrm{Mingxing Chen}}
\affiliation{Key Laboratory for Matter Microstructure and Function of Hunan Province, Key Laboratory of Low-Dimensional Quantum Structures and Quantum Control of Ministry of Education, School of Physics and Electronics, Hunan Normal University, Changsha 410081, China.}
	\author{Chao Tang}
\thanks{Authors to whom correspondence should be addressed: Chaoyu He, \textcolor{blue}{chaoyu@xtu.edu.cn}; Tao Ouyang, \textcolor{blue}{ouyangtao@xtu.edu.cn}; Chao Tang, \textcolor{blue}{tang$\_$chao@xtu.edu.cn.} }
\affiliation{Hunan Key Laboratory for Micro-Nano Energy Materials and Device and School of Physics and Optoelectronics, Xiangtan University, Xiangtan 411105, Hunan, China}

	\begin{abstract}
Ferroelectricity has a wide range of applications in functional electronics and is extremely important for the development of next-generation information storage technology, but it is difficult to achieve due to its special symmetry requirements. In this letter, based on van der Waals stacking, a generic model is proposed for realizing ferroelectric devices, where a freely movable center layer is packaged in two fixed and symmetrically stacked layers. In this model, the ferroelectric phase transition can be realized between the two equivalent and eccentric ground stacking-states with opposite polarizations. 
By means of first-principles calculations, taking the $ h $-BN/$ h $-BN/$ h $-BN and $ h $-BN/Graphene/$ h $-BN as feasible models, we carefully evaluate the magnitude of ferroelectricity. The corresponding polarizations are estimated as 1.83 and 1.35 pC/m, respectively, which are comparable to the sliding ferroelectricity. Such a new tri-layer model of vertical ferroelectricity can be constructed by arbitrary van der Waals semiconducting materials, and usually holds low switching barrier. Optimized material combinations with remarkable polarization are highly expectable to be discovered from the huge candidate set for future information storage.

	\end{abstract}
\footnotetext{$\ast~$Both authors contributed equally.}
	\maketitle

	\thispagestyle{fancy}
	\lhead{\ifthenelse{\value{page}=1}{}{\thepage\ of \pageref{LastPage}}}
	\rhead{\ifthenelse{\value{page}=1}{}{\small\leftmark}}
	\fancyhead[C]{\ifthenelse{\value{page}=1}{ }{}}
	\renewcommand{\headrulewidth}{0.8pt}
	\renewcommand{\footrulewidth}{0.0pt}

\par Spontaneous polarization caused by non-centrosymmetry  is a key character of ferroelectric material. The ferroelectric polarizations can be switched in a reversible approach by an external electric field, which plays a critical role in the applications of electronic devices, sensors, field-effect transistors, and memory devices.\cite{RN1518,RN1519,RN1185,RN1455} Since the introduction of the first ferroelectric memory chip in 1993, ferroelectric memory has been regarded as a strong competitor for the next generation of advanced memory devices. \cite{RN1520} Extensive research has focused on three-dimensional ferroelectric materials, such as perovskite structures including BaTiO$_{3}$, PbTiO$_{3}$, BiFeO$_{3}$, and Pb[Zr$_{x}$Ti$_{1-x}$]O$_{3}$. \cite{RN1521,RN1522,RN1523,RN1185} However, conventional ferroelectricity usually diminish at nanoscale due to the depolarization field, hindering miniaturization and preventing ferroelectric memory from keeping pace with integrated circuits. For instance, the widely-used lead zirconate titanate (PZT) exhibit obvious degradation in performance below 70 nm thickness \cite{RN1520}. As a result, commercially available ferroelectric memory has stagnated at the 130 nm process node, which limiting the development of high-density, low-cost, and highly reliable devices. 
\par In recent years, two-dimensional ferroelectric materials have captured widespread attentions in views of their advantages of thinner thickness, low energy consumption, and ease of miniaturization. \cite{RN1185,RN1184} In 2016, Liu et al. \cite{RN1309} reported reversible out-of-plane ferroelectricity in 4 nm-thick CuInP$_{2}$S$_{6}$ with a Curie temperature (T$_{c}$) of 320 K. Additionally, Chang et al. \cite{RN1239} discovered stable in-plane polarization in SnTe (T$_{c}$ = 270 K) at the monolayer limit. In 2017, Ding et al.\cite{RN1525} have also predicted robust ferroelectricity in a 2D III$_{2}$-VI$_{3}$ compound (In$_{2}$Se$_{3}$) at the monolayer limit. These findings indicate the potential to overcome the bottleneck of miniaturization in ferroelectric memory devices. However, no matter three-dimensional and two-dimensional ferroelectric materials, both of them require specific crystal structures. Such feature makes the intrinsic two-dimensional ferroelectric materials exceedingly scarce. Most of the two-dimensional materials are centrosymmetry
 and nonpolar structures. It is interesting that Wu et al \cite{RN1310} have proposed that centrosymmetry non-ferroelectric monolayers, under certain stacking configurations, can undergo symmetry breaking in their bilayer form, leading to the emergence of vertical ferroelectricity. Moreover, the sliding of atomic layers within nonpolar van der Waals structures under the influence of an electric field could give rise to a transfer of interlayer charge and reversal of spontaneous polarization as well.  This phenomenon known as sliding ferroelectricity, which diverges from the traditional ionic displacement-based ferroelectric switching mechanism. Subsequently, sliding ferroelectricity has been experimentally confirmed in various two-dimensional material systems, including bilayer$ h $-BN, $\gamma$-InSe, and MoS$_{2}$, by Pablo Jarillo-Herrero’s team and others researchers. \cite{RN1526,RN1527,RN1529,RN1530,RN1531}
\setParDis
\par However, for micro- and nanoscale memory devices, such interlayer sliding can introduce lattice distortions, such as in-plane stretching, compression, or folding, which may compromise the ferroelectric properties of the device over prolonged usage. To overcome these drawbacks, an A/B/A type tri-layer model is proposed in this work for realizing van der Waals ferroelectricity by vertical moving, where the sliding is prohibited. By using the first-principles calculations, the feasibility of our A/B/A type model for vertical ferroelectric device is carefully investigated based on the $ h $-BN/$ h $-BN/$ h $-BN and $ h $-BN/Graphene/$ h $-BN structures.  The results indicate that there exits significant polarization magnitudes in both $ h $-BN/$ h $-BN/$ h $-BN and $ h $-BN/Graphene/$ h $-BN, offering potential applications in next-generation non-volatile memory devices. Our model can be constructed by arbitrary van der Waals semiconducting materials, suggesting numerous practical material combinations can be considered to realize such vertical ferroelectric device with remarkable polarizations.

\par First-principles calculations based on density functional theory (DFT) are performed using the Vienna Ab-initio Simulation Package (VASP) code\cite{RN1124}, employing the generalized gradient approximation\cite{RN1125} (GGA) with the Perdew-Burke-Ernzerhof (PBE)\cite{RN1126} functional. The projector augmented-wave (PAW) method is used to describe\cite{RN1127,RN1126} the interactions between ions and valence electrons. A plane-wave cutoff energy is set as 500 eV and Monkhorst-Pack\cite{RN1123} $\Gamma$-centered k-point sampling is used for the first Brillouin zone, with VASPKIT\cite{RN851} controlling k-point grid spacing in reciprocal space within 0.02 $\times$ 2$\pi$\AA$^{-1}$.  Convergence criteria for energy and forces during optimization are 10$^{-10}$ eV and 10$^{-4}$ eV/\AA, respectively. A vacuum layer with a minimum thickness of 15 \AA $ $ is included to avoid periodic mirror interactions along the $z$ direction. In all geometry optimizations, the Grimme-D2\cite{RN1542} method is used for van der Waals correction.  Ferroelectric polarization is evaluated using Berry’s Phase method\cite{RN968,RN1131}, with dipole corrections considered along the $z$ direction. The barrier for ferroelectric switching is determined using the climbing image nudged elastic band (CINEB) method\cite{RN143}.


\begin{table}[b!] 
	\small

	\caption{ Energy Variation of $ h $-BN/$ h $-BN/$ h $-BN and $ h $-BN/Graphene/$ h $-BN Devices under Different Stacking Configurations (in eV/atom).}
	\label{tbl:tbl-1}
	
	
	\begin{tabular*}{\hsize}{@{}@{\extracolsep{\fill}}lccccc@{}}

		\hline
		\multirow{2}{*}{Device} & \multicolumn{5}{c}{Stacking type}  \\
		\cline{2-6} 
		& I	& II  &III& IV	&V\\
		\hline
		$ h $-BN/$ h $-BN/$ h $-BN & -8.890 & -8.884 & -8.889 & -8.884 & -8.891\\
		$ h $-BN/Graphene/$ h $-BN & -9.016 & -9.017 & -9.021 & & \\
		\hline
	\end{tabular*} 
\end{table}

\begin{figure}[!t] 
	\centering
	\includegraphics[width=1.0\linewidth]{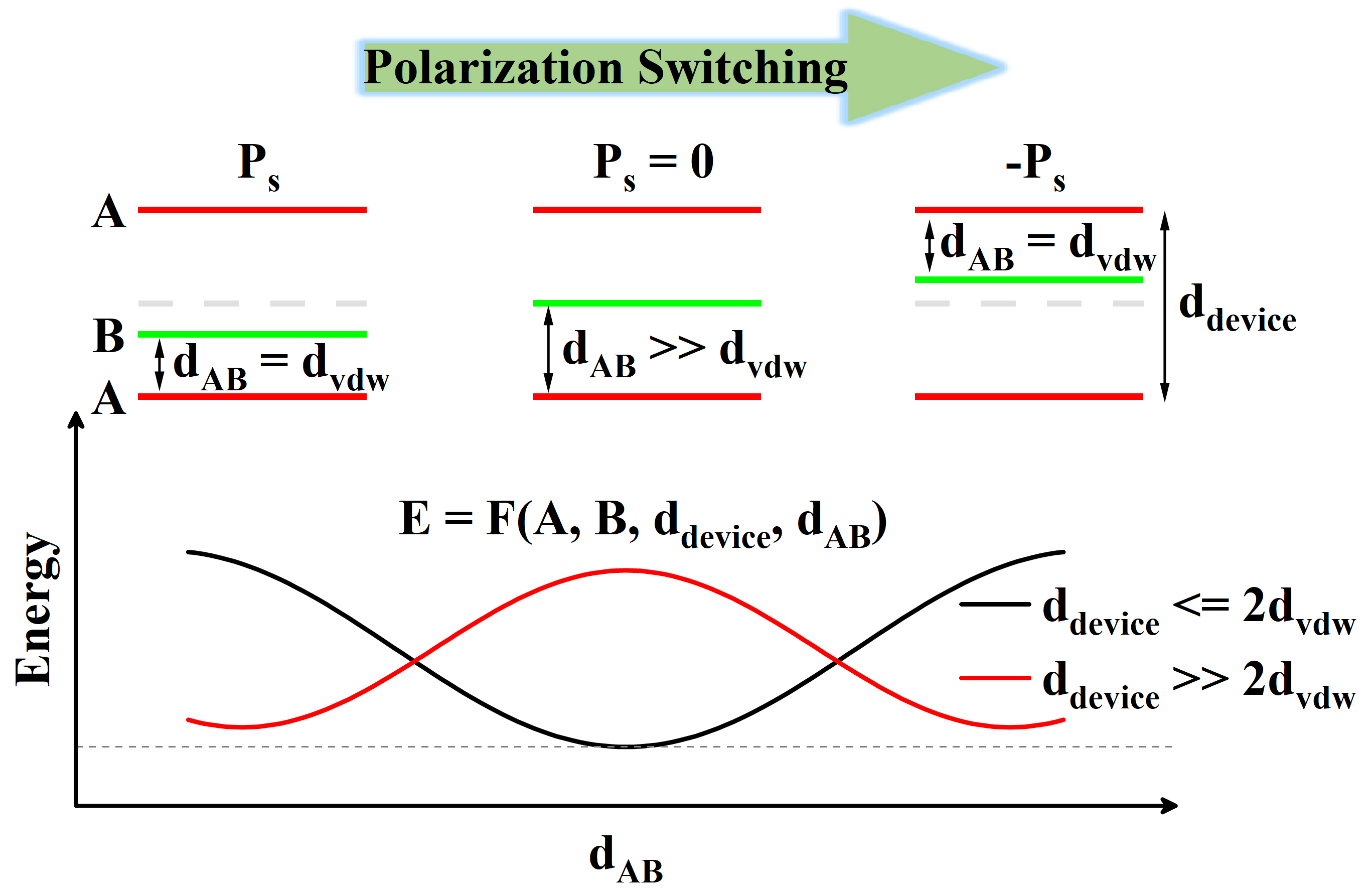}
	\caption{Schematic diagram of the conceptual model for a van der Waals vertical ferroelectric device.}
	\label{fgr:fig-1}
\end{figure}
\begin{figure}[t!] 
	\centering
	\includegraphics[width=1.0\linewidth]{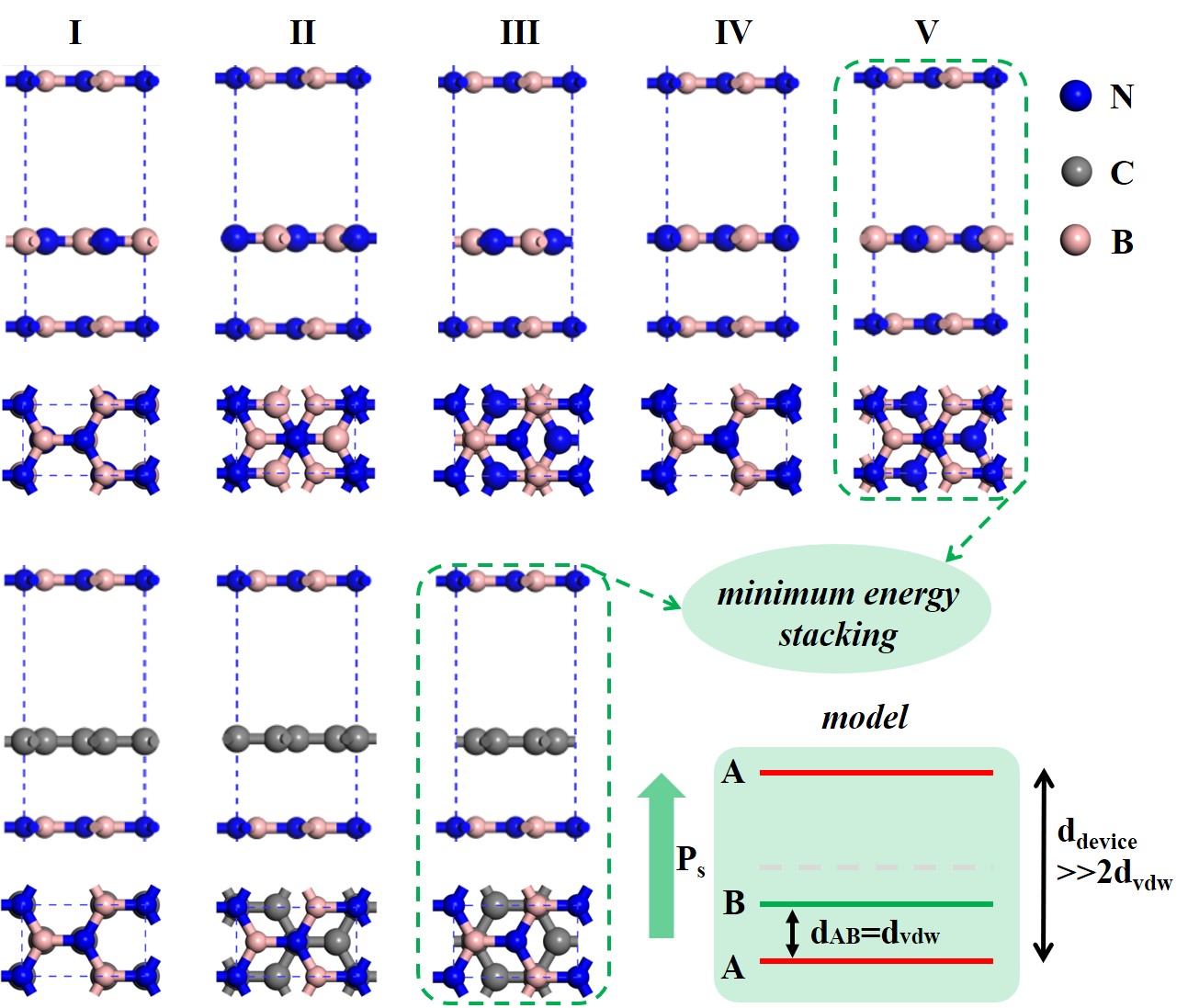}
	\caption{The $ h $-BN/$ h $-BN/$ h $-BN and $ h $-BN/Graphene/$ h $-BN tri-layer model is studied in various stacking configurations.}
	\label{fgr:fig-2}
\end{figure}
\begin{figure}[!t] 
	\centering
	\includegraphics[width=1.0\linewidth]{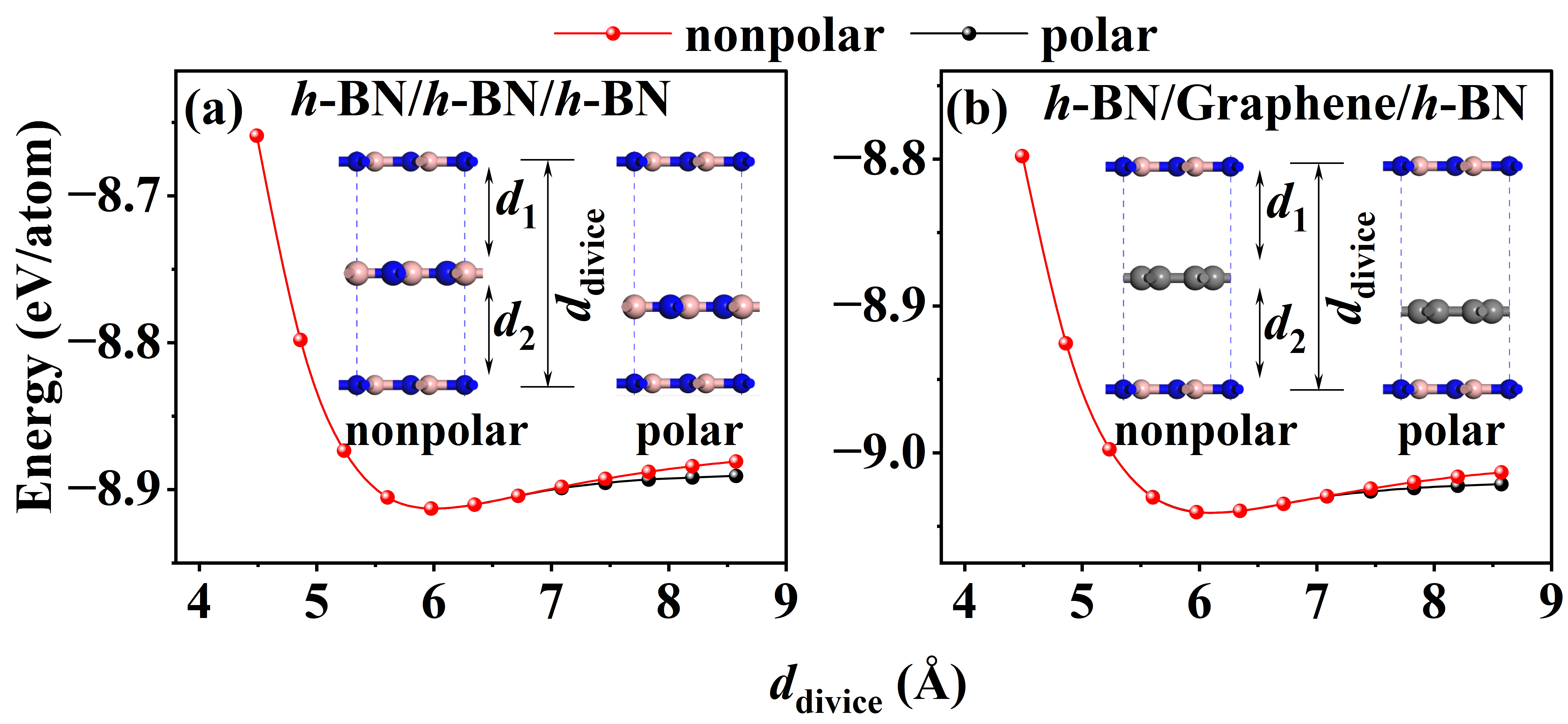}
	\caption{Variation of the energy of $ h $-BN/$ h $-BN/$ h $-BN and $ h $-BN/Graphene/$ h $-BN layers as a function of distance from the device.}
	\label{fgr:fig-3}
\end{figure}
\begin{figure*}[t!] 
	\centering
	\includegraphics[width=1\linewidth]{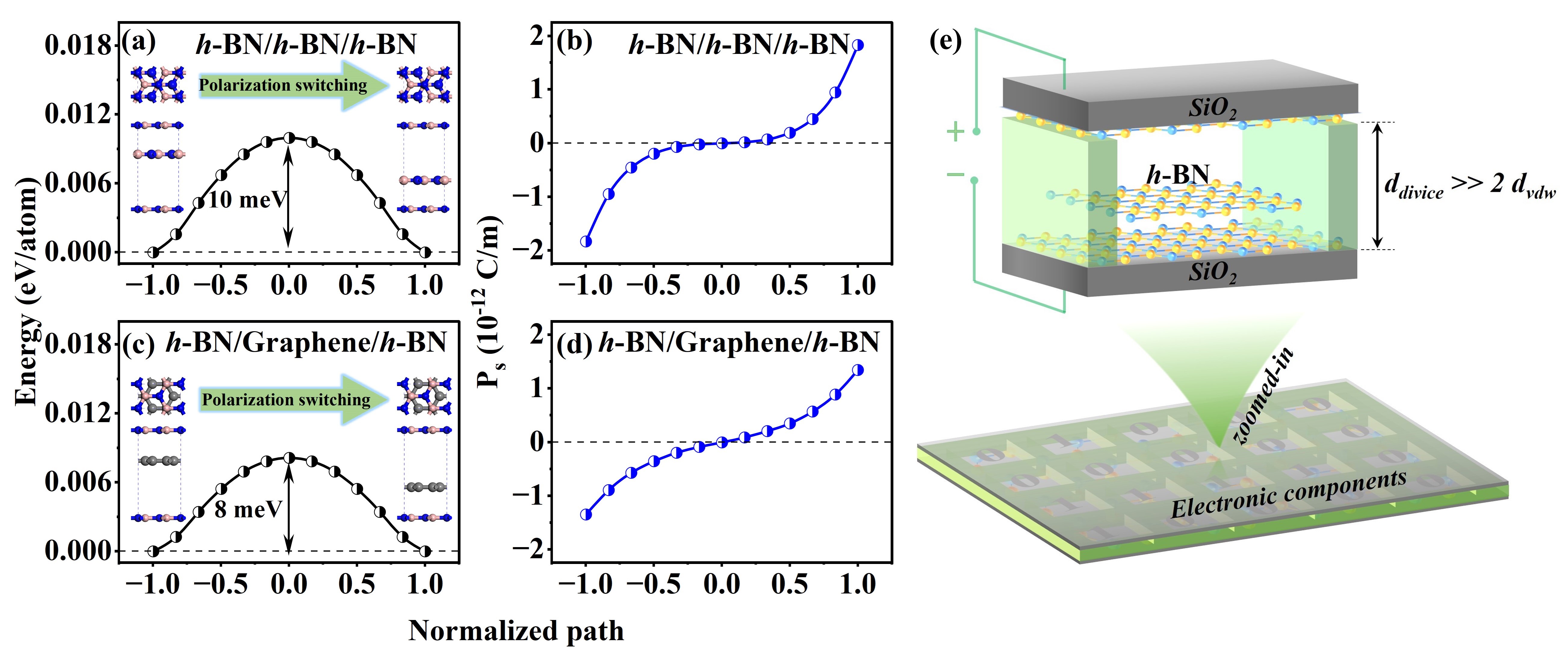}
	\caption{(a)-(b) Ferroelectric switching energy barriers and spontaneous polarization of the ferroelectric in $ h $-BN/$ h $-BN/$ h $-BN and $ h $-BN/Graphene/$ h $-BN devices. (e) Prototype device based on van der Waals vertical ferroelectricity.}
	\label{fgr:fig-4}
\end{figure*}
\par Van der Waals materials typically exhibit a layered structure, consisting of multiple atomic or molecular layers stacked together, with stability maintained by Van der Waals forces among the layers. In Fig. \ref{fgr:fig-1}, we present an A/B/A type tri-layer model for realizing ferroelectric device via using layered Van der Waals materials. Such an A/B/A type model always exhibit out-of-plane vertical polarization if the ground stacking-state is eccentric. It is easy to know that the energy of this model can be expressed as a function of A, B, $d_{device}$, and $d_{AB}$, represented as
\begin{equation}
	E=F\left(A, B, d_{\text {device }}, d_{A B}\right)
\end{equation}
Here, A and B represent different material types or different stacking arrangements of the same material. The distance between the device layers, $d_{device}$, and the distance between layer A and layer B, $d_{AB}$, are important parameters in this model. For different values of $d_{device}$ and $d_{AB}$, the energy curve of this model exhibits two distinct states. When the distance between the device layers is smaller than or equal to twice the Van der Waals distance between A and B ($d_{device}$ $\leq$ $2d_{vdw}$), layer B prefers to be optimized to the center of the system to form a centric ground stacking-state. In this case, the energy curve exhibits an upward-opening parabolic shape as indicated in the black solid line in Fig. \ref{fgr:fig-1}. The device is centrosymmetric and does not exhibit any polarization. When the distance between the device layers is obviously greater than twice of the common Van der Waals distance between A and B ($d_{device}$ $\gg$ $2d_{vdw}$), however, layer B tends to be on one side of layer A due to the Van der Waals interactions and the principle of minimum energy. In such an eccentric ground stacking-state case, the energy curve, shown as the red curve in Fig. \ref{fgr:fig-1}, exhibits a typical double-well shape. Such a non-centrosymmetric device always exhibits spontaneous polarization and it possesses two equivalent and eccentric ground stacking-states. As layer B translate between the two equivalent eccentric ground stacking-states through vertical moving, the polarization reverses from $P_{s}$ to $-P_{s}$. Clearly, by artificially controlling $d_{device}$ $\gg$ $d_{AB}$, we can realize a vertically ferroelectric device with the lowest energy state and out-of-plane polarization.
\par In order to investigate the validity of the proposed A/B/A model of vertical ferroelectricity on realizing out-of-plane polarization, the well-known single layer materials $ h $-BN and graphene are taken as research platform.
 Two types of vertically ferroelectric devices, $ h $-BN/$ h $-BN/$ h $-BN and $ h $-BN/Graphene/$ h $-BN, are constructed, as illustrated in Fig. \ref{fgr:fig-2}. In the $ h $-BN/$ h $-BN/$ h $-BN device, there are five stacking configurations labeled as I, II, III, IV, and V.
 For the $ h $-BN/Graphene/$ h $-BN device, it only exists three stacking configurations: I, II, and III. This is mainly because of the enhanced symmetry from the intermediate graphene layer, where the stacking configurations IV and V degenerate with I and II. 
The lowest energy stacking configurations for the two devices are confirmed to be V and III, respectively, and their respective energies are listed in Table I. The competition between the polarization state and non-polarization state under different device values is crucial in this model. It determines whether the device exhibits spontaneous polarization and maintains long-term stability. Therefore, we perform a study to investigate this competition for the two lowest energy stacking configurations under different device values. As depicted in Fig. \ref{fgr:fig-3}, when the device distance ($d_{device}$) is less than 7 Å, the polarization state is not stable. Similar to the behavior shown in Fig. 1, when $d_{device}$ is less than or equal to twice the van der Waals distance ($2d_{vdw}$), the non-polarization state dominates. The polarization state, initially located on the slope of the potential energy surface, converts to the non-polarization state during geometric optimization. However, when $d_{device}$ increases to 7 \AA, the polarization state becomes energetically favorable. Under the influence of van der Waals forces, layer B is biased towards one side of layer A, resulting in the spontaneous polarization of the entire system. This further validates the accuracy of our previous model.
\par The magnitude of polarization is a crucial characteristic influencing the occurrence and properties of ferroelectric phenomena. Figures. \ref{fgr:fig-4}(b) and (d) illustrate the evaluation of spontaneous polarization magnitude in $ h $-BN/$ h $-BN/$ h $-BN and $ h $-BN/Graphene/$ h $-BN van der Waals ferroelectric devices by using the Berry’s Phase\cite{RN968,RN1131} method. The results indicate that the $ h $-BN/$ h $-BN/$ h $-BN system exhibits a larger spontaneous polarization (1.83 pC/m) compared to the $ h $-BN/Graphene/$ h $-BN system (1.35 pC/m). This discrepancy can be attributed to the pronounced electronegativity differences among the B, C, and N elements (B: 2.04, C: 2.55, N: 3.04). The interconnected $ h $-BN/$ h $-BN/$ h $-BN structure is characterized by B-N-B interactions between layers through van der Waals forces, while the $ h $-BN/Graphene/$ h $-BN structure relies on B-C-B interactions, leading to smaller electronegativity differences and smaller spontaneous polarization. 
\par The ferroelectric polarization switching barrier represents the energy necessary to overcome the atomic rearrangements involved in reversing the polarization direction of a ferroelectric material under the influence of an external electric field. Using the CINEB method, we calculate the polarization switching barriers for the $ h $-BN/$ h $-BN/$ h $-BN and $ h $-BN/Graphene/$ h $-BN systems as 10 and 8 meV/atom, respectively (as shown in Figs. \ref{fgr:fig-4}(b) and (d)). These results suggest that the polarization reversal is likely to occur in both van der Waals ferroelectric devices under specific external fields. A low polarization switching barrier indicates easier control of the polarization direction during the application process. Meanwhile, it can significantly reduce the cost of polarization reversal, which could give rise to an enhanced device performance, such as faster response rates and higher sensitivity.
\par The discovery of van der Waals vertical ferroelectricity has expanded the potential for achieving switchable ferroelectricity in a diverse range of materials, rather than be restricted to specific ones. The recent breakthroughs in fabrication techniques for sandwiched graphene and trilayer magic angle graphene have opened up possibilities for harnessing this property\cite{RN1534,RN1539}. These advancements present new opportunities for manufacturing more efficient electronic devices and components.  In Fig. \ref{fgr:fig-4}(e), we showcase the implementation of switchable ferroelectricity by utilizing the tri-layer A/B/A device model. With the help of a vertical electric field, the movement of the intermediate $ h $-BN layer can be control in this model, facilitating upward and downward displacements, and further enabling the switching between 0 and 1 in logical circuits.

\par In this work, a conceptual prototype with vertical ferroelectricity based on the van der Waals materials has been proposed. The first-principles calculations also demonstrate the feasibility of assembling such devices using the van der Waals materials and the capability to exhibit switchable vertical ferroelectricity. The results demonstrate that the van der Waals ferroelectric devices based on$ h $-BN/$ h $-BN/$ h $-BN and $ h $-BN/Graphene/$ h $-BN have measurable polarization magnitudes of 1.83 and 1.35 pC/m, respectively. These magnitudes are comparable to the sliding ferroelectric polarization observed in conventional crystal systems. This study overcomes the limitations imposed by the symmetry of traditional ferroelectric materials, and providing novel insights and directions for the development of ferroelectric-based storage technologies. Moreover, it should be emphasized that such ferroelectric devices fabricated based on this model hosting numerous advantages, including low switching barriers, strong controllability, and rapid response, foreshowing the promising prospects for future advancements and applications.

\par This work is supported by the National Natural Science Foundation of China (Grant Nos. 11974299 and 11974300), the Scientific Research Foundation of Education Bureau of Hunan Province (Grant Nos. 20B582, 20K127, 19C1746, and 20A503), the Youth Science and Technology Talent Project of Hunan Province (Grant No. 2022RC1197), the Science Fund for Distinguished Young Scholars of Hunan Province of China (Grant No. 2021JJ10036), and the Program for Changjiang Scholars and Innovative Research Team in University (Grant No. IRT13093).

\section*{AUTHOR DECLARATIONS} 
\subsection*{Conflict of Interest} 
\par The authors have no conflicts to disclose. 
\subsection*{Author Contributions}
\textbf{Yuwen Zhang:} Writing - original draft, Visualization. 
\textbf{Chunfeng Cui:} Writing - original draft. 
\textbf{Chaoyu He:} Formal analysis, editing, Funding acquisition.
\textbf{Tao Ouyang:} Formal analysis,Writing- review $\&$ editing, Funding acquisition.
\textbf{Mingxing Chen:} Software, Formal analysis.
\textbf{Jin Li:} Formal analysis, Funding acquisition.
\textbf{Chao Tang:} Writing- review $\&$ editing, Project administration, Funding acquisition.
\subsection*{DATA AVAILABILITY}
\par The data that support the findings of this study are available from the corresponding author upon reasonable request.
\bibliographystyle{apsrev4-2}
\bibliography{reference.bib}

\bibliographystyle{apsrev4-2}
\bibliography{reference}

\end{document}